\documentclass[11pt,a4paper]{emulateapj}
\usepackage{natbib}
\usepackage{amssymb, amsmath, amsbsy}
\usepackage{color}
\usepackage[linkcolor=blue]{hyperref}%

\usepackage{epsfig}
\usepackage{amsmath}
\usepackage{amssymb}
\usepackage{natbib}
\usepackage{subfigure}
\usepackage{graphicx}

\slugcomment{submitted to {\it The Astrophysical Journal Letter}}

\newcommand{\hi}{H {\sc i} }
\newcommand{\civ}{C {\sc iv}}
\newcommand{\heii}{He {\sc ii}}

\begin{document}
\def\mean#1{\left< #1 \right>}
\newcommand{\sbunit} {
  erg\ s$^{-1}$ cm$^{-2}$ arcsec$^{-2}$}
  
\title{Keck/Palomar Cosmic Web Imagers (KCWI/PCWI)  Reveal an Enormous Ly$\alpha$ Nebula in an Extremely Overdense QSO Pair 
Field at $z=2.45$ }
\author{Zheng Cai\altaffilmark{1,6}, Erika Hamden\altaffilmark{2}, Matt Matuszewski\altaffilmark{2}, J. Xavier Prochaska\altaffilmark{1}, Qiong Li\altaffilmark{3},  
  Sebastiano Cantalupo\altaffilmark{4}, Fabrizio Arrigoni Battaia\altaffilmark{5}, Christopher Martin \altaffilmark{2},  James D. Neill\altaffilmark{2}, Donal O'Sullivan\altaffilmark{2}, 
  Ran Wang \altaffilmark{3}, Anna Moore  \altaffilmark{2}, Patrick Morrissey\altaffilmark{2} }
\affil{$^1$ UCO/Lick Observatory, University of California, 1156 High Street, Santa Cruz, CA 95064, USA }
\affil{$^2$ Cahill Center for Astrophysics, California Institute of Technology, 1216 East California Boulevard, Mail code 278-17, Pasadena, California 91125, USA}
\affil{$^3$ Kavli Institute for Astronomy and Astrophysics, Peking University, Beijing 100871, People's Republic of China}
\affil{$^4$ Department of Physics, ETH Zurich, CH-8093,  Zurich, Switzerland}
\affil{$^5$ European Southern Observatory, Karl-Schwarzschild-Str. 2, D-85748 Garching bei M\"unchen, Germany}
\affil{$^{6}$ Hubble Fellow}





\begin{abstract}
Enormous Ly$\alpha$ nebulae (ELANe) represent the extrema of Ly$\alpha$ nebulosities. They have detected extents of $>200$ kpc in Ly$\alpha$ and Ly$\alpha$ luminosities $>10^{44}$ erg s$^{-1}$. 
The ELAN population is an ideal laboratory to study the interactions between galaxies and the intergalactic/circumgalactic medium (IGM/CGM) given their brightness and sizes.
The current sample size of ELANe is still very small, and the few $z\approx2$ ELANe discovered to date are all associated with local overdensities of active galactic nuclei (AGNs). 
Inspired by these results, we have initiated a survey of ELANe associated with QSO pairs using the Palomar and Keck Cosmic Web Imagers (PCWI/KCWI). In this 
letter, we present our first result: the discovery of ELAN0101+0201 associated with a QSO pair at $z=2.45$. 
Our PCWI discovery data shows that, above a 2-$\sigma$ surface brightness of $1.2\times10^{-17}$ \sbunit, the end-to-end size of ELAN0101+0201 is $\gtrsim 232$ kpc. 
We have conducted follow-up observations using KCWI, resolving multiple 
Ly$\alpha$ emitting sources within the rectangular field-of-view of $\approx 130\times165$ projected kpc$^2$, and obtaining their 
 emission line profiles at high signal-to-noise ratios.  Combining both KCWI and PCWI, our observations confirm that ELAN0101+0201 resides in an extremely overdense environment.  
 Our observations further support that a large amount of cool ($T\sim10^4$~K) gas could exist in massive halos (M~$\gtrsim10^{13}$~M$_\odot$) at $z\approx2$. 
 Future observations on a larger sample of similar systems will provide statistics of how cool  gas is distributed in massive 
  overdensities at high-redshift  and strongly constrain the evolution of the intracluster medium (ICM).
\end{abstract}





\section{Introduction}\label{sec:introduction}

Recent observations have lead to the discovery of the first sample of enormous Ly$\alpha$ nebulae (ELANe) at $z=2-3$ \citep{cantalupo14, hennawi15, cai17, arrigoni18}. 
 These ELANe have a Ly$\alpha$ size exceeding the diameters of their massive dark matter halos and their Ly$\alpha$ luminosities are greater than $10^{44}$ erg s$^{-1}$.  
Their morphology and kinematics are ideal properties to constrain models of  galaxy formation and galaxy-IGM interactions \citep{martin15, arrigoni18}.

Although Ly$\alpha$ nebulae with sizes of 100 kpc are commonly found around luminous QSOs at $z\gtrsim3$ \citep{borisova16}, two out of 17 nebulae in this sample have sizes $>200$ kpc which meet the ELANe definition.  
Further, at $z\approx 2$, QSO-Ly$\alpha$ nebulae appear to be rarer. Arrigoni Battaia et al. (2016) conducted deep narrowband images on 15 $z\approx2.2$ QSOs and did not detect bright nebulae with SB $\approx$ 10$^{-17}$ \sbunit\ at distances $>50$ kpc around any of these QSOs. These observations suggest that typical $z\approx2$ QSOs may be surrounded by gas $\sim$100 times less dense than that responsible for ELANe \citep{hennawi13, cantalupo17}.

All  $z\approx2$ ELANe found to date reside in local AGN overdensities: the Slug and MAMMOTH-1 nebulae \citep{cantalupo14, cai17} contain two AGNs. The Jackpot nebula \citep{hennawi15} 
is associated with an AGN quartet. There are two possible reasons for these detections. A stronger UV radiation field may boost the extended  Ly$\alpha$ emission into the detectable regime \citep{cantalupo05, kollmeier10}, and 
a higher CGM density may also yield brighter Ly$\alpha$ fluorescence \cite[e.g.,][]{cantalupo12}. 
Thus, AGN pairs or groups may be ideal sites to search for ELANe, because they provide both the strong meta-galactic ionizing fluxes and gas overdensities required for detections. This has motivated our survey using QSO pairs. In addition, recent studies have suggested that SDSS QSO pairs trace protoclusters (\citealt{onoue18}), and thus the extended Ly$\alpha$, \civ, \heii\ lines traced by QSO pairs can be used to probe cool ($T\sim10^4$ K)  and warm ($T\sim10^5$ K) gas in the early ICM \citep{valentino16}. 
 
Guided by the above observational and theoretical insights, we have conducted a survey using the Palomar and Keck Cosmic Web Imagers (PCWI/KCWI) to search for ELANe associated with QSO pairs at $z\approx 2$. Our goals are to significantly enlarge the number of confirmed ELANe at $z\approx 2$, and then systematically understand the gas budget and kinematics throughout the dark matter halos of these objects. Our results will be compared to nebulae powered by isolated QSOs at $z\approx 2-3$ and further used to constrain the ICM evolution. 

In this letter, we present our first discovery in the survey: ELAN0101+0201, associated with the QSO pair at $z=2.4$ using both PCWI and KCWI observations. The projected separation of these two QSOs is $\approx 10"$ (82.5 physical projected kpc) and the redshift separation is about 1000 km s$^{-1}$. The brighter QSO (Q1) has an AB magnitude of $i=18.1$ and the fainter QSO (Q2) has $i=21.6$ ($\approx$30 times fainter than Q1).  The statistical results on the full sample of 12 sources will be presented in upcoming papers. In \S2, we introduce our observations and data reductions. 
In \S3, we provide the discovery of the ELAN and the gas kinematics. In \S4, we provide a discussion of the results. 
We assume a $\rm{\Lambda}$CDM cosmology with $\Omega_m= 0.3$, $\Omega_{\Lambda}=0.7$ and $h=0.70$.

\section{Observations}\label{sec:observations}

\subsection{Sample Selection}\label{sec:Target}

We searched for QSO pairs from the SDSS-IV QSO database which contains $\sim$ 200,000 QSO spectra at $2.0<z<2.5$ \citep{paris17}. 
Our sample contains 12 QSO pairs, where each consists of two QSOs with 2D separation smaller than $1'$, and velocity offset $ < 2,000$ km s$^{-1}$. Further, we require 
at least one of them to have $g<19$. 
The observations are conducted in two stages. Stage 1 is a shallow PCWI program to search for extended emission; 
stage 2 is a deeper targeted search with KCWI to conduct high resolution, high signal-to-noise observations to study 
the gas kinematics and emission lines in greater detail. 
ELAN0101+0201, at $z\approx2.4$, is our first target from this sample with both PCWI and follow-up KCWI data.


\subsection{PCWI and KCWI Observations, Data Reduction}

The PCWI observations of ELAN0101+0201 were carried out on UT-20161130 and 
UT-20171215. The total exposure time is 4-hours on-source and 4-hours on the blank sky. Individual exposures were acquired using the standard CWI nod-and-shuffle technique \citep{martin14}. 
PCWI uses a 40$"\times 60"$ reflective image slicer with 24 $40" \times 2.5"$ slices. We put the brighter QSO (Q1) at the IFU center. We used the Richardson grating ($R=2000$) and 
an Asahi/blue filter. 
We use the standard CWI pipeline to reduce the data \citep{martin14}. 

The KCWI observations of ELAN0101+0201 were carried out on UT-20171021 and UT-20171119. 
The BM1 grating and medium slicer were chosen. This setting yields a field-of-view (FoV) of 16.8$"\times 20"$, and a spatial resolution of $\approx 0.7"$ along the slicer and 
seeing-limited sampling ($\approx 1.0"$) perpendicular to the slicer. The spectral resolution is $R=4000$. 
The datacube has wavelength coverage of $\lambda= 3950-4800$ \AA, centered on the QSO Ly$\alpha$ redshift ($\lambda \approx 4200$ \AA). The total on-source exposure time is 120 minutes, in four 
10 minute and four 20 minute individual exposures. 
To maximizing the observing efficiency, we used the ``offset-target-field" to construct the sky datacube, rather than nod-and-shuffle.
The offset-target has a different redshift and is within two degrees of ELAN0101+0201. We obtained the sky of each wavelength channel using the median value 
after masking the point-source of the offset-target. 
The KCWI pipeline\footnote{KCWI pipeline: https://github.com/kcwidev/kderp/blob/master} was used to reduce the data. 
For each image, we subtracted the bias, corrected the pixel-to-pixel variation, removed cosmic-rays, 
corrected the geometric transformation, and 
perform the wavelength calibration. 
For each channel, we subtract the sky determined by the offset-target field.  Flux calibration was performed using the standard star BD+28D4211 at the beginning of the night. The final datacube is combined from the inverse variance weight of individual exposures. In addition, we use CubeExtractor package to further perform the high order flat-field corrections and to remove the point spread function (PSF) of the bright QSO (Q1) (Cantalupo in 
prep; see also \citet{borisova16} for a description of main routines).

\section{Results}

\subsection{The Discovery of an Enormous Ly$\alpha$ Nebula}

In the upper left panel of Figure~1, we present the PCWI PSF-subtracted Ly$\alpha$ narrowband image. The 4-hour on-source integration with nod-and-shuffle mode yields a 1-$\sigma_{\rm{SB}}$ of $4.5\times 10^{-18}$ erg s$^{-1}$ cm$^{-2}$ arcsec$^{-2}$ over a velocity range of $\pm$500 km s$^{-1}$ around the systematic redshift 
of Q1 ($z=2.4510$). 
To a 2-$\sigma_{\rm{SB}}$ contour, the ELAN has an extended size of 27.9$"$, corresponding to a physical size of 232.2 kpc. The total Ly$\alpha$ luminosity of ELAN0101+0201 is 
$L_{\rm{Ly\alpha}}= 4.5\times10^{44}$ erg s$^{-1}$.

In the upper right panel of Figure~1, we present the KCWI observations. The size of the ELAN exceeds the KCWI FoV (the orange box). The 2-hour integration yields a flux density of 1-$\sigma\approx 2.4\times 10^{-19}$ erg s$^{-1}$ cm$^{-2}$ \AA$^{-1}$, corresponding to a 1-$\sigma_{\rm{SB}}$ of $1.0\times 10^{-18}$ erg s$^{-1}$ cm$^{-2}$ arcsec$^{-2}$ over a velocity range of $\pm500$ km s$^{-1}$ around the systematic redshift 
of Q1, a factor of $\approx5\times$ deeper than PCWI observations. KCWI has a much higher spatial resolution for resolving the sub-structures in the Ly$\alpha$ nebula. Both PCWI and KCWI detect extended Ly$\alpha$ emission and a projected filament that appears to connect both QSOs (yellow asterisks in Fig.~1).

\subsection{The kinematics of the Ly$\alpha$ Nebula}

From the KCWI observations, we derive a flux-weighted velocity map (Figure~1, lower left). 
The zero-point of the velocity is set at the systematic redshift of Q1 ($z=2.4510$), determined from the MgII emission in the SDSS spectrum. 
From the velocity map, 
there are no clear velocity structures (e.g., rotation or bi-polar outflow signatures). 
The majority of the nebula has a velocity offset within $\pm150$ km s$^{-1}$ from the zero-point. 
A high velocity component with $v\gtrsim 150$ km s$^{-1}$ is seen in the North-West, and this component corresponds to a LAE discussed below. 
The velocity dispersion map (Figure 1, bottom right) is obtained using the second moment of the flux distribution. 
The majority of the nebula has a velocity dispersion $\lesssim 250$ km s$^{-1}$, indicating a Ly$\alpha$ line width 
of $\lesssim 587$ km s$^{-1}$ throughout most of the nebular region covered by KCWI.

\subsection{Overdensity of Ly$\alpha$ Emitters}

In Figure~2, we show six 1-D spectra from the KCWI datacube, including Q1 ($i=18.1$), Q2 ($i=21.6$), and a KCWI resolved bright LAE (LAE1). The positions labeled f1, f2 and f3 
are bright nodes in the nebular region resolved by KCWI.

LAE1 has a higher velocity than the nebular redshift (Fig.~1). Figure~2 shows that LAE1 has a highly asymmetric 
Ly$\alpha$ profile. The Ly$\alpha$ emission can be fit using one broad (thick orange line) 
component with the FWHM of 840 km s$^{-1}$ and one narrow component (thin orange line) 
with the FWHM of 170 km s$^{-1}$. The left wing of the broad 
component is strongly absorbed. The possible causes of this absorption are discussed in \S4.3.

In Figure~3, we show four 1-D spectra from the PCWI datacube using a $2"\times3"$ aperture. All sources have $>5\sigma$ detection of the emission lines. These sources reside outside the KCWI FoV. Assuming all these emission lines are Ly$\alpha$,  their
Ly$\alpha$ luminosities are $L_{\rm{Ly\alpha}}= 1.6-3.2\times10^{42}$ erg s$^{-1}$ ($L^*_{\rm{Ly\alpha}}=2.1\times10^{42}$ erg s$^{-1}$,  \citealt{ciardullo12}). Including 
LAE1, we detect four LAE candidates and two QSOs in an $40"\times60"$ area, within a redshift range of $ z = 2.44$ -- 2.46 (Table~1). 


\subsection{The Pair of Ly$\alpha$ Absorbers}

In the spectra of Q1 and Q2, we detect strong Ly$\alpha$ absorbers at 
 -710 km s$^{-1}$ reference to QSO Ly$\alpha$ emission ($z=2.441$) (gray vertical lines in Figure~2). 
 Using a Markov-Chain Monte Carlo (MCMC) analysis (Figure~4), we found that the column density of both absorbers have $N_{\rm{HI}}\approx 10^{15-18}$ cm$^{-2}$ in $>95.5\%$ (Fig.~4). These absorbers are not associated with any Ly$\alpha$-emitting galaxies in the KCWI observed field within a velocity offset of $\pm1000$ km s$^{-1}$ around the absorption redshift. Within the KCWI FoV, we put a 5-$\sigma$ upper limit of $0.1\ L^{*}_{\rm{Ly\alpha}}$ for the counterpart galaxy within $R \approx50$ projected kpc around Q1.  
 This  corresponds to a Ly$\alpha$-based SFR of $\lesssim0.3$ M$_\odot$ yr$^{-1}$ \citep{kennicutt12}. Within the PCWI FoV, we put a 5-$\sigma$ upper limit of $L_{\rm{Ly\alpha}}= 0.6\ L^*_{\rm{Ly\alpha}}$ within $R\lesssim 150$ projected kpc, corresponding to a Ly$\alpha$-based SFR of $\lesssim1.5$ M$_\odot$ yr$^{-1}$. 
 

\section{Discussion}

We have identified ELAN0101+0201 with an end-to-end size of $\gtrsim232$ kpc.  
Such a large and luminous nebula is rarely found, especially around $z\approx2$ QSOs. 
The filament is elongated along the projected line connecting the two QSOs. With KCWI observations, 
we further resolved a LAE (LAE1 in Fig.2) which has a highly asymmetric Ly$\alpha$ emission. 

\subsection{The Enormous Nebula at $z\approx 2$ and its Powering Mechanism}

\citet{borisova16} show that $z\gtrsim3.1$ QSOs are generally associated with nebulae having projected sizes larger than 100 physical kpc. 
However, at $z\approx2$, few isolated ultraluminous QSOs appear to power nebulae \citep{arrigoni16}. 
ELAN0101+0201 is associated with a $z\approx2$ QSO pair. To understand the physical properties, we investigate the powering mechanism of this nebula. A few processes can power the extended Ly$\alpha$ emission: photoionization, 
 shock-heated gas by an AGN outflow, 
and resonant scattering from the QSO broad-line-region (Cantalupo 2017 for a review).

The kinematics analysis disfavors that a strong outflow plays a major role in powering this nebula. 
The velocity map of ELAN0101+0201 (Fig.1) does not show obvious bipolar outflow signatures (Cai et al. 2017b). 
\citet{alexander10} and \citet{harrison14} also suggest that AGN outflows have a high-velocity of $v_{\rm{max}}\gtrsim$ 1500 km s$^{-1}$ which is not the case for ELAN0101+0201. 
The velocity pattern is similar to nebulae found by Borisova et al. (2016) at $z\approx3$. 
The velocity dispersion measurement (\S3.2 and Figure~1) is consistent with a nebula illuminated by a radio-quiet system (FWHM $<700$ km s$^{-1}$) \citep{villar07, borisova16}.

We now consider the photoionization scenario. Since Q2 is expected to be $30\times$ fainter than Q1 and most nebular emission is around Q1, we 
assume that the nebula is photoionized by Q1. 
We consider both optically thin and thick CGM photoionized by Q1.  

Following \citet{hennawi13}, the surface brightness of an 
optically thin cloud due to recombination can be expressed as follows: 
\begin{equation}
\begin{split}
SB_{\rm{Ly\alpha}}^{\rm{thin}}&= 8.8\times10^{-20} \Big( \frac{1+z}{3.253}\Big)^{-4}\times \\ 
& \Big(\frac{f_{\rm{C}}^{\rm{thin}}}{0.5}\Big) \Big(\frac{N_{\rm{H}}}{10^{20.5}\ \rm{cm^{-2}}}\Big)\  \rm{erg}\ \rm{s}^{-1} \ \rm{cm}^{-2}\ \rm{arcsec}^{-2}
\end{split}
\end{equation}
Arrigoni Battaia et al. (2015a,b) and Hennawi et al. (2015) presented 
that the smooth Ly$\alpha$ morphology suggest the covering factor 
$f^{\rm{thin}}_{\rm{C}}\ge 0.5$. Here  
we assume the $f^{\rm{thin}}_{\rm{C}}=0.5$. 
Using a statistical sample, 
\citet{lau16} find that ${N_{\rm{H}}}\approx 10^{20.5\pm 1.0}$ cm$^{-2}$
within 200 kpc in the QSO halo. 
We thus assume $N_{\rm{H}} =10^{20.5}$ cm$^{-2}$ in Eq.(1). 
Given the average SB$_{\rm{Ly\alpha}}\approx 1\times 10^{-17}$ erg s$^{-1}$ cm$^{-2}$ arcsec$^{-2}$ at $R\approx 50$ kpc from Q1, 
the hydrogen number density is $n_{\rm{H}}\approx 1$ cm$^{-3}$. This value 
is typical for the interstellar medium, but not typical for 
CGM where simulations predict $n_{\rm{H}}\sim 10^{-3}$ cm$^{-3}$ \cite[e.g.,][]{shen13}. 
The ELAN0101+0201 has a CGM density 2--3 orders of magnitude higher than the statistical QSO sample at $z\sim2$ (Arrigoni Battaia et al. 2016), 
suggesting that this QSO pair may contain an unusually high-density CGM compared to typical isolated QSOs at $z\sim2$. 

If we consider the optically thick scenario, \citet{hennawi13} suggest that the SB can be estimated as follows: 
\begin{equation}
\begin{split}
SB_{\rm{Ly\alpha}}^{\rm{thick}} = & 5.3 \times10^{-17} \Big(\frac{1+z}{3.45}\Big)^{-4} 
\Big(\frac{f_{\rm{C}}^{\rm{thick}}}{0.5}\Big) \Big(\frac{R}{160\ \rm{kpc}}\Big)  \times \\ 
& \Big (\frac{L_{\rm{\nu_{\rm{LL}}}}}{10^{30.9}\rm{erg} \rm{s}^{-1} \rm{Hz}^{-1} } \Big) 
 \rm{erg}\ \rm{s}^{-1} \ \rm{cm}^{-2}\ \rm{arcsec}^{-2}
 \end{split}
\end{equation}
The $i$-band magnitude of the bright QSO is 18.1, and the inferred Log $\frac{L_{\rm{\nu_{\rm{LL}}}}}{\rm{erg}\ \rm{s^{-1}}}= 31.1$, 
obtained from scaling the composite QSO spectrum to match the $i$-band magnitude of Q1 \citep{arrigoni15a, lusso15}. 
Assuming $f_{\rm{C}}^{\rm{thick}}=0.5$, 
the predicted SB$_{\rm{Ly\alpha}}\sim 10^{-15}$ erg s$^{-1}$ cm$^{-2}$ at $\approx 50$ kpc from Q1, 
two orders of magnitude higher than the observed SB. Thus, the UV-illuminated 
optically thick gas may have a small covering factor, or the illumination due to optically thick gas is 
not the dominant powering mechanism.

\subsection{The Association of Cool Gas and 
 Overdense Environements at $z\approx2$}

Using Hyper-Suprime-Cam survey and SDSS DR12 QSO database,
 \citet{onoue18} found that 
QSO pairs trace $> 5$-$\sigma$ overdense regions 
over 15 co-Mpc. These overdense regions could 
be protoclusters with $M_{\rm{z=0}}> 10^{14}$ M$_\odot$. 
 ELAN0101+0201 is associated with one of the smallest-separation QSO pairs at $z\approx2$. 
 In the KCWI FoV ($\sim100$ physical kpc), 
 we revealed another bright LAE with a broad Ly$\alpha$ emission. If this 
 LAE is another AGN, then the AGN clustering in this field is even stronger than the AGN quartet 
 system reported in Hennawi et al. (2015), implying a physical connection between this ELAN 
 and a massive overdensity. 
 

From Figure~3, over a larger FoV of $40"\times60"$, we detect another three 
  bright LAE candidates from the PCWI data with $L_{\rm{Ly\alpha}}>0.8\times L_{\rm{Ly\alpha}}^*$ (Table~1). 
    In random fields, one expects $\approx 0.05$ LAEs 
within this luminosity range and survey volume at $z\approx2.2$ 
\cite[e.g.,][]{ciardullo12} 
  This further suggests that ELAN0201+0101 resides in the progenitor of a massive cluster.

 From \S4.1, to interpret the discovery of ELAN, we assume the optically-thin, cool gas ($T\sim10^4$~K)  that 
 has an average column density of $N_{\rm{H}}\approx 10^{20.5}$ cm$^{-2}$ 
 across the ELAN. The cool gas mass can be calculated using the $N_{\rm{H}}\times A_{\rm{nebula}}\times m_{\rm{H}}$, 
 where $A_{\rm{nebula}}$ is the area of the nebula, and $m_{\rm{H}}$ is the atomic hydrogen mass. 
 One can calculate that this ELAN could contain a reservoir of cool gas with the mass of $\approx 10^{11}$ M$_\odot$ 
 (also see \citealt{hennawi15}). 
The association between cool gas reservoirs and massive overdensities is unexpected. Cosmological simulations suggest that by $z \sim 2$--3, 
baryons in the cluster progenitors ($M \gtrsim 10^{13}$ M$_\odot$) 
 are dominated by a hot, shock-heated gas with $T \sim 10^7$ K \citep{fumagalli14, faucher15}. 
  ELAN0101+0201 provides further challenges to these simulations. 
Recently, \citet{mccourt18} and \citet{ji18} suggest that such inconsistencies may be 
solved if the cold gas takes the form of tiny fragments, sparsely 
distributed analogous to a mist.


\subsection{The Profile of the Ly$\alpha$ Emission}

With KCWI, we reveal high S/N Ly$\alpha$ emission and absorption throughout this system. 
From Figure~2, LAE1 has a highly asymmetric 
Ly$\alpha$ profile, and contains a broad component 
(FWHM=840 km s$^{-1}$). 
The broad component is much wider than typical star-forming galaxies 
($\approx 300$--400 km s$^{-1}$, e.g., \citealt{trainor16}), indicating that LAE1 
could be an AGN, making this field likely hosting three AGNs in a $\sim100$ kpc scale.  
The left wing of the broad component is strongly suppressed by the absorption.

Ly$\alpha$ is a resonant line and its profile may 
depend on the kinematics of the local interstellar medium (ISM). %
An ISM outflow could cause the asymmetric Ly$\alpha$ emission \citep{dijkstra06, yang14}.
An alternative source of asymmetry is enhanced IGM absorption 
along the sightline that could suppress the blue peak, making the Ly$\alpha$ line asymmetric. 
From \S4.2, ELAN0101+0201 resides in a massive overdensity. 
Such overdensities may be associated with a significant IGM \hi overdensity \citep{cai16}. 
This \hi reservoir is likely to absorb 
the blue wing of the LAE1 and make the Ly$\alpha$ emission to be highly asymmetric. 


In the future, we will present our complete sample of 12 QSO pairs, using PCWI and KCWI observations.  
Through deep IFS observations on this unique sample, we will conduct a statistical study of the Ly$\alpha$ 
spatial extent and gas kinematics, revealing how CGM gas fuel massive halos. 
With this survey, we will build a unique statistical sample to study the
multi-phase gas content in overdensities at $z\sim2$ and 
provide important constraints to the 
formation and evolution of the hot, shock-heated
intracluster medium (ICM) observed in local Universe.

{\bf Acknowledgement:} ZC acknowledges the supports provided by NASA through
the Hubble Fellowship grant HST-HF2-51370 awarded by the
Space Telescope Science Institute, which is operated by the
Association of Universities for Research in Astronomy, Inc., for
NASA, under contract NAS 5-26555. EH acknowledges the National Science Foundation under Grant No. AST-1402206. 
SC gratefully acknowledges support from 
Swiss National Science Foundation grant PP00P2\_163824. 

\begin{figure*}
\centering
\includegraphics[width = \linewidth]{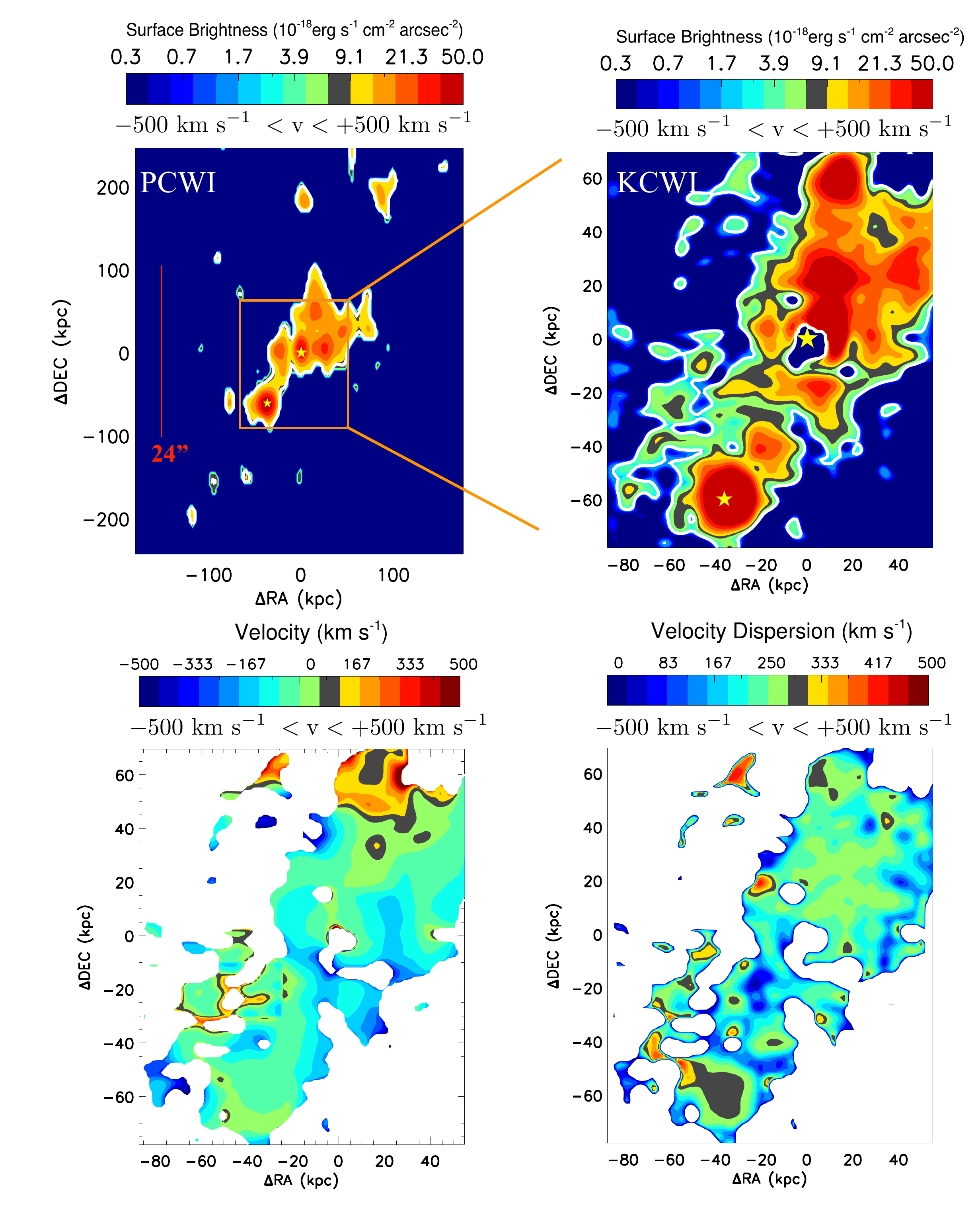}
\caption{The upper left panel shows the point-spread-function (PSF) subtracted narrowband image of the ELAN0101+0201 observed using the Palomar/CWI (PCWI). 
From PCWI observations, this Ly$\alpha$ nebula has an end-to-end size of 232 kpc. 
With 8-hour PCWI nod-and-shuffle observations, the 1-$\sigma$ surface brightness (SB) limit is $4.5\times 10^{-18}$ erg s$^{-1}$ cm$^{-2}$ arcsec$^{-2}$.
The white contour presents the 2-$\sigma$ SB level.  
The upper right panel shows the PSF-subtracted KCWI high-resolution Ly$\alpha$ image. The KCWI field of view (FoV) is marked by the orange box in the left panel. 
The 1-$\sigma$ surface brightness limit is $\approx 1 \times10^{-18}$ erg s$^{-1}$ cm$^{-2}$ arcsec$^{-2}$ (white curve indicates the 3-$\sigma$ SB contour). 
The lower two panels show the flux-weighted velocity and the flux-weighted velocity dispersion map using KCWI. }
\label{figure1}
\end{figure*}

\begin{figure*}
\centering
\includegraphics[width = \linewidth]{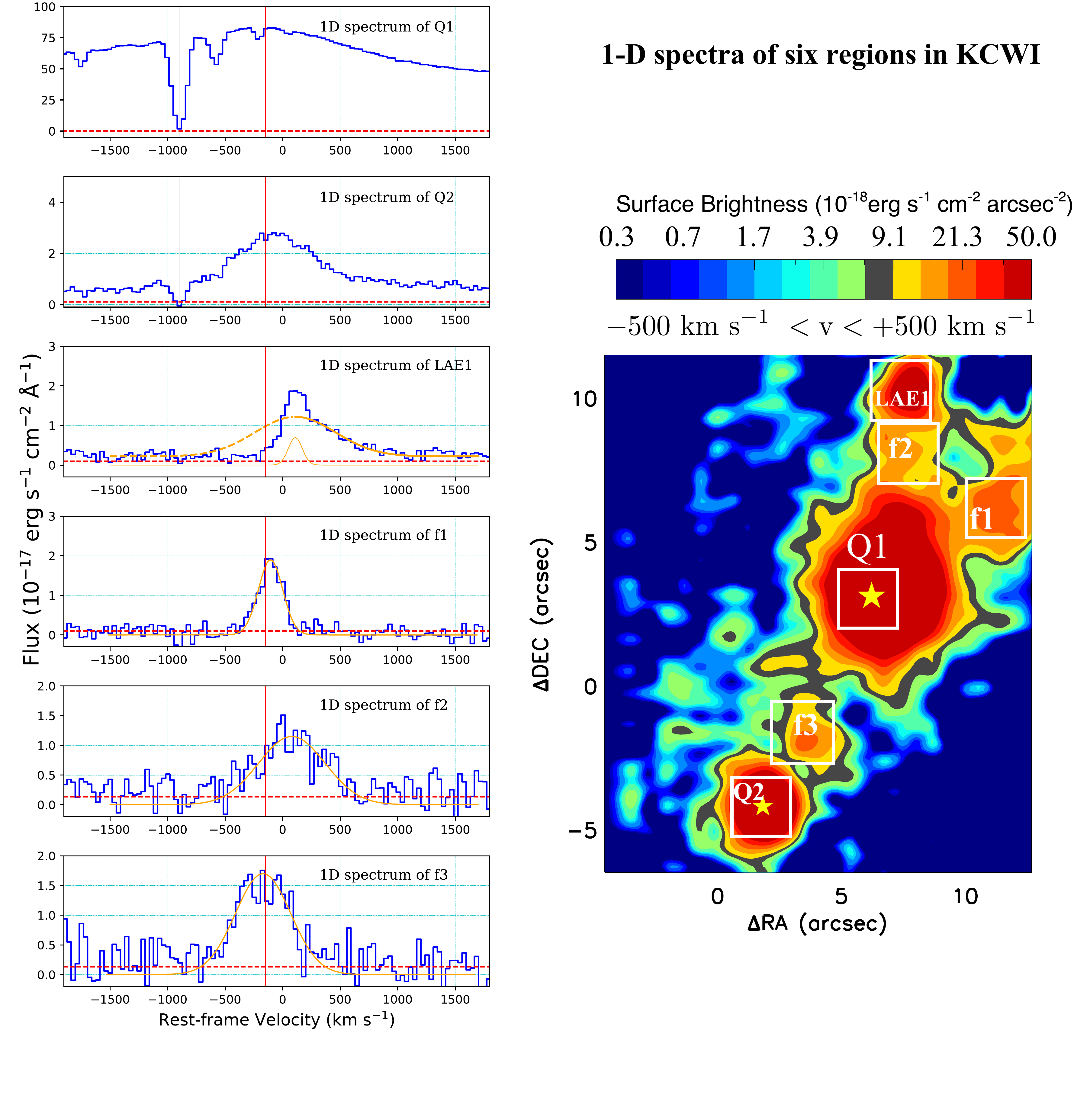}
\caption{The 1-D spectra (blue line) extracted from the KCWI datacube with the noise overplotted (red dotted line). 
The zeropoint is set using the systematic redshift of Q1. The vertical red lines indicate the Ly$\alpha$ redshift of Q1. 
We extract six regions: brighter QSO (Q1), fainter QSO (Q2), one Ly$\alpha$ emitter (LAE1), and three brighter nodes in 
the nebular region (f1, f2, and f3). We mark strong Ly$\alpha$ absorption detected in the spectra of Q1 and Q2 using 
gray vertical lines. For the Ly$\alpha$ emission of LAE1, we use one broad (thick orange line) with the FWHM of 840 km s$^{-1}$ 
and one narrow component (thin orange line) with the FWHM of 170 km s$^{-1}$. The blue wing of the broad component 
is significantly absorbed (dotted yellow line). The Ly$\alpha$ emission of f1, f2, and f3 can be fit with a single Gaussian.}
\label{figure2}
\end{figure*}

\begin{figure*}
\centering
\includegraphics[width = \linewidth]{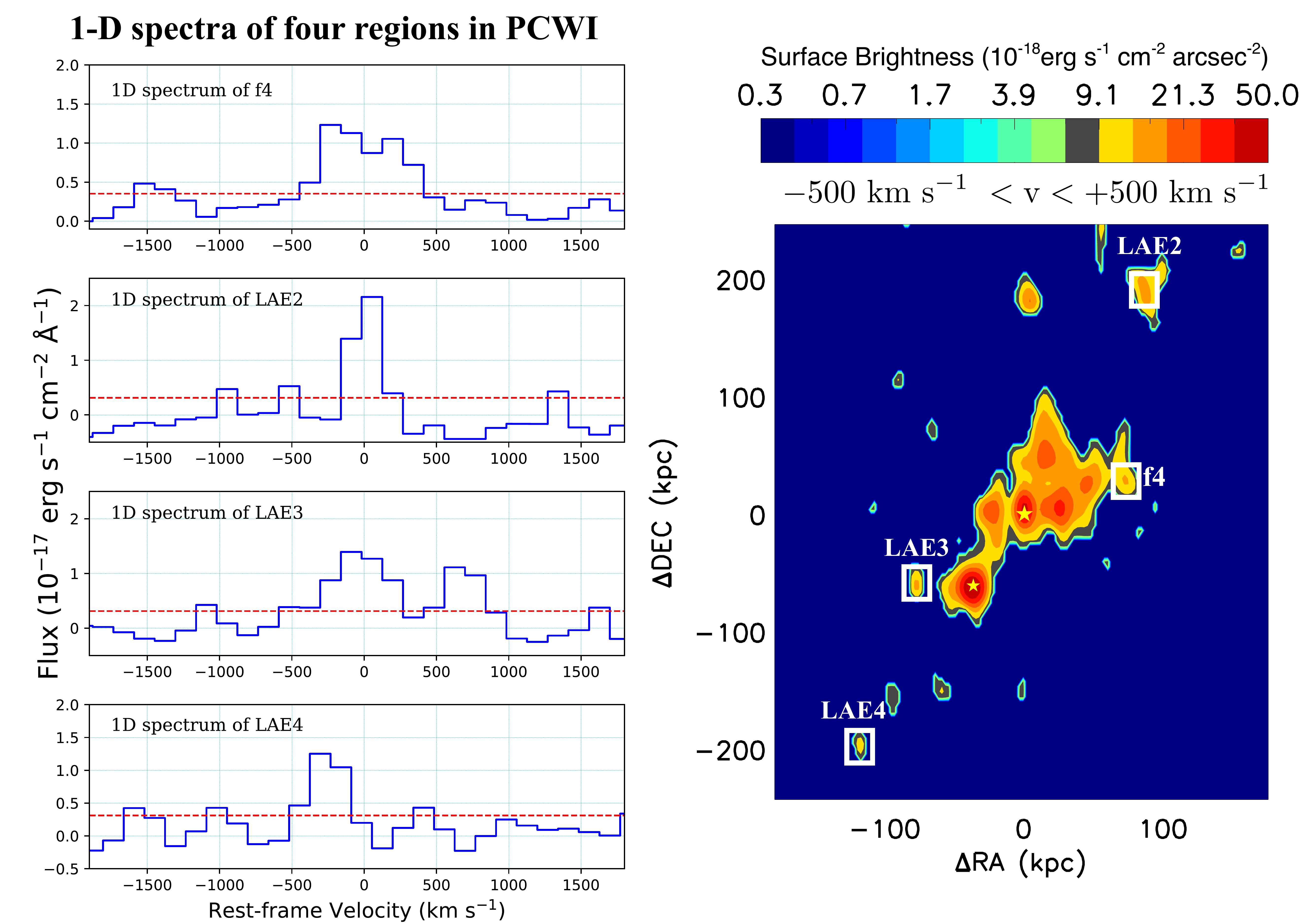}
\caption{The 1-D spectra (blue line) extracted from the PCWI datacube with the noise overplotted (red). 
We extract four sources outside the KCWI field of view (white rectangles). LAE2, LAE3 and LAE4 are three 
LAE candidates with the emission lines $> 5\sigma$ over $2" \times 3"$ apertures. 
We extract the 1-D spectrum of a nebular region (f4). Including LAE1 resolved by KCWI (Figure 2), we demonstrate 
that ELAN0101+0201 is embedded in a significantly overdense field (see details in \S4.2).}
\label{figure2}
\end{figure*}

\begin{figure*}
\centering
\includegraphics[width = \linewidth]{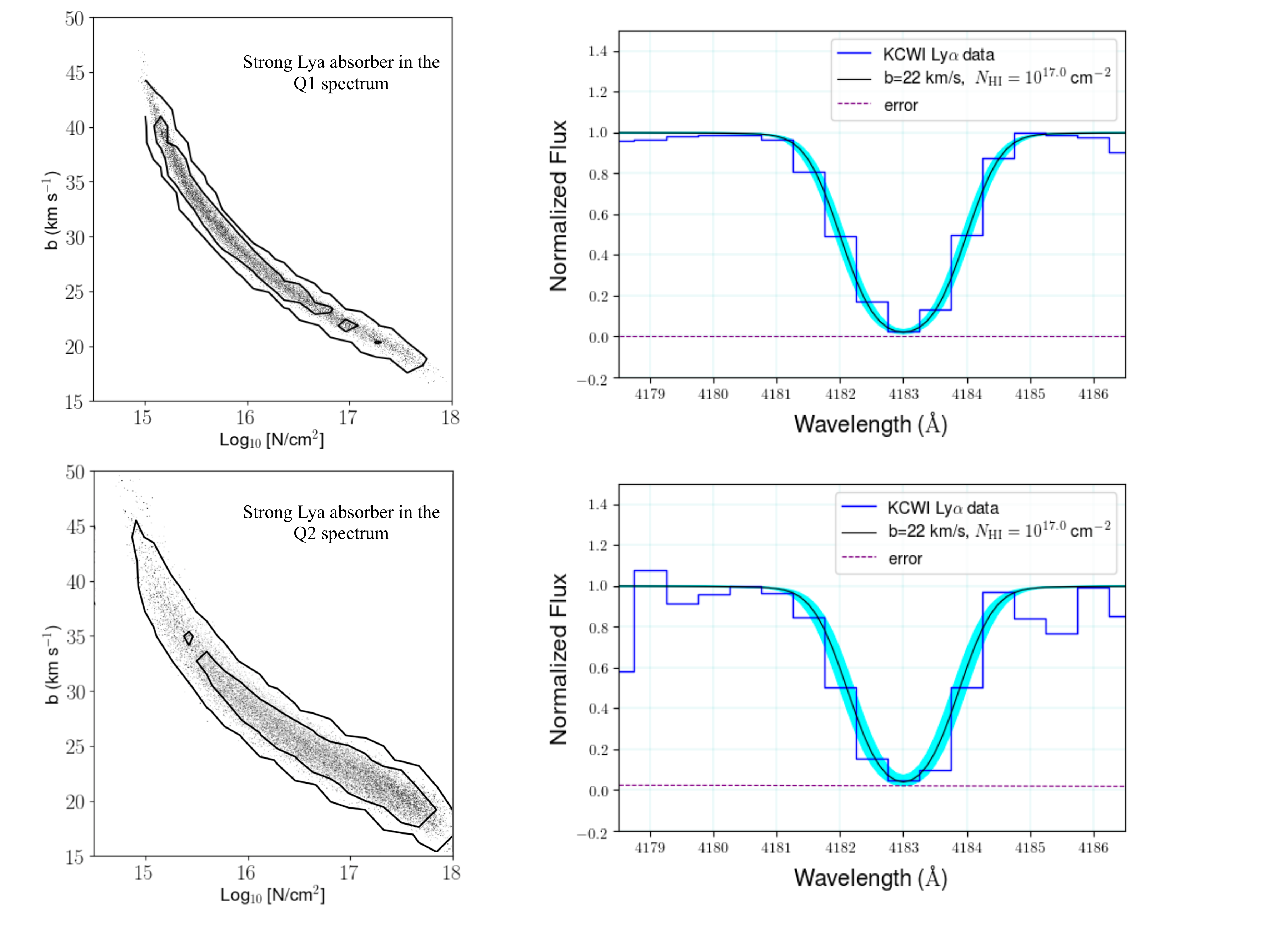}
\caption{The Markov Chain Monte Carlo (MCMC) analysis of the column density ($N_{\rm{HI}}$) and Doppler parameter ($b$) 
of the strong absorbers in the spectrum of Q1 (upper) and Q2 (lower). The absorption system is marked by vertical gray 
lines in Figure 2. In the right panel, we provide the 1-D absorber fit using $N_{\rm{HI}} = 10^{17}$ cm$^{-2}$ and $b = 22$ km s$^{-1}$, with 1-$\sigma$ uncertainty 
using a cyan region. 
We did not detect galaxy 
counterparts associated with this absorption group from current data (see details in \S3.4).  }
\end{figure*}

\begin{table*}[!h]
\caption{Sources in the field of ELAN0101+0201 } 
\label{table:PopIII_SFR}	
\centering 
\begin{tabular}{| c | c | c | c | c | c |} 
\hline\hline 
Name & RA & DEC  &  $z$  & $L_{\rm{Ly\alpha}}$  &  IFS\\ 
 &   &    &     &  (10$^{42}$ erg s$^{-1}$)  & \\ 

\hline
Q1 (brighter QSO) & 01:01:16.54   &  +02:01:57.4   &  2.4510 (Mg II) & $802.0$ & KCWI \\
 \hline
Q2 (fainter QSO) & 01:01:16.85   &  +02:01:49.8   &  2.459 (Ly$\alpha$) &  $10.1$ & KCWI\\
\hline
LAE1 (could be an AGN) &   01:01:16.44    &    +02:02:04.5  & 2.462 (Ly$\alpha$)  & $2.90\pm0.03$ & KCWI\\ 
\hline
\hline
LAE2  & 01:01:16.30  &  +02:02:25.2  & 2.449 (Ly$\alpha$) & $2.8\pm0.2$  & PCWI\\
\hline
LAE3  & 01:01:16.45  &  +02:01:50.0  & 2.453 (Ly$\alpha$) & $3.1\pm0.2$ & PCWI\\
\hline
LAE4  & 01:01:17.34  &  +02:01:38.5  & 2.446 (Ly$\alpha$) & $1.6\pm0.2$ & PCWI\\
\hline
\end{tabular}
\end{table*}


\end{document}